\documentclass[nofootinbib, aps, prd, a4paper, 10pt, superscriptaddress, eqsecnum, showkeys]{revtex4-2}
\usepackage[a4paper, top=2cm, bottom=2cm, left=2.5cm, right=2.5cm]{geometry}

\usepackage{amsmath}
\usepackage{amsfonts}
\usepackage{amsthm}
\usepackage{bm}
\usepackage{mathrsfs}
\usepackage{braket}

\usepackage{graphicx}
\usepackage{booktabs}

\usepackage{xcolor}
\usepackage[pdfusetitle]{hyperref}
\hypersetup{colorlinks = true, allcolors = blue}

\usepackage{orcidlink}

\begin{document}

\title{Dark Energy in Ghost-free non-local Gravity}
\author{S.D. Odintsov}
\email{odintsov@ieec.cat} \affiliation{ICREA, Passeig Luis
Companys, 23, 08010 Barcelona, Spain}
\affiliation{Institute of
Space Sciences (IEEC-CSIC) C. Can Magrans s/n, 08193 Barcelona,
Spain}
 \author{V.K. Oikonomou}
\email{v.k.oikonomou1979@gmail.com;voikonomou@gapps.auth.gr}
\affiliation{Physics Department, Observatory, Aristotle University
of Thessaloniki, Thessaloniki, Greece} \affiliation{Center for
Theoretical Physics, Khazar University, 41 Mehseti Str., Baku,
AZ-1096, Azerbaijan}
\author{G.S. Sharov}
 \email{sharov.gs@tversu.ru}
 \affiliation{Tver state university, Sadovyj per. 35, 170002 Tver, Russia}
 \affiliation{International Laboratory for Theoretical Cosmology,
Tomsk State University of Control Systems and Radioelectronics (TUSUR), 634050 Tomsk,
Russia}

\begin{abstract}
Ghost-free non-local gravity is investigated with regards to its
late-time dynamics. Viable solutions in this model are confronted
with the observational data including the Pantheon+ catalogue of
Type Ia supernovae, the Dark Energy Spectroscopic Instrument, the
measurements of baryon acoustic oscillations and the Hubble
parameter estimations $H(z)$. The ghost-free non-local gravity is
found to be successful in these tests in comparison to the
$\Lambda$CDM model and can be also comparable with the generalized
exponential $F(R)$ gravity scenario. However the model encounters
difficulties when the data from the above observations and the
cosmic microwave background radiation data are combined. In tests
with the whole set of Pantheon+, DESI, $H(z)$ and CMB data, the
generalized exponential $F(R)$ model is essentially more
successful. This success is related with the dynamical behavior of
its effective dark energy equation of state evolving from a
phantom to a quintessence phase during the late-time epoch,
whereas the ghost-free non-local model demonstrates only a
quintessence behavior. Hence the ghost-free non-local gravity
scenario is successful only when the Pantheon+, DESI and $H(z)$
data are considered. The generalized exponential $F(R)$ model
satisfies the viability conditions and in tests with all
observational data including CMB surpasses the $\Lambda$CDM model
in $\chi^2$ statistics and also with information criteria.
\end{abstract}


\maketitle

\section{Introduction}

During the last two years, one can see groundbreaking changes in
observational cosmology based on the new observations of Baryon
Acoustic Oscillations (BAO) from the Dark Energy Spectroscopic
Instrument (DESI) \cite{DESI:2024,DESI:2025zgx} along with the
Type Ia Supernovae (SNe Ia) data  from the Pantheon+, Union3,
DES5Y catalogues \cite{PantheonP:2022,Union3:2023,DES5Y:2024}.
These new observational data can shed light on the nature and
properties of dark energy that drives the late-time accelerated
expansion of the Universe observed during last three decades. This
acceleration and dark energy as its origin, is successfully
described in the frameworks of the $\Lambda$-Cold-Dark-Matter
model ($\Lambda$CDM) with the cosmological constant $\Lambda$
playing the role of dark energy, with the equation of state (EoS)
for dark energy is constant: $P_{DE}/\rho_{DE}=-1$
\cite{Peebles:2003}.

The $\Lambda$CDM model during last decades encountered several
problems in theory and observations, including vague physical
nature of its dark components, the coincidence problem of their
densities nowadays, the fine-tuning for $\Lambda$, the Hubble
constant tension and other questions about dark matter and dark
energy induced numerous alternative cosmological scenarios
including modifications of General Relativity (see reviews
\cite{Peebles:2003,BambaCNO:2012,reviews2,reviews3}). Cosmologists
suggested different approaches  for solving the problems related
with the dark energy nature and structure and several scenarios
explaining the Hubble constant tension between early-time
estimations of $H_0$ from Cosmic Microwave Background radiation
(CMB) \cite{Planck:2018}, and local distance-ladder measurements
by SH0ES collaboration were proposed
\cite{Riess2021,DiValentino:2020naf,
DiValentino:2020evt,Krishnan:2020obg,Krishnan:2021dyb,OdintsovSGStens:2021,Vagnozzi:2021quy,Vagnozzi:2021tjv,Ye:2021iwa,Ferlito:2022mok,Lee:2022cyh,Adil:2023ara,
Hogas:2023pjz,Menci:2024rbq,DiValentino:2025sru}.

The strongest challenge for the $\Lambda$CDM model appeared two
years ago and was further confirmed in 2025 with new BAO DESI data
\cite{DESI:2024,DESI:2025zgx}. These data contradicted constant
EoS for dark energy corresponding to the $\Lambda$CDM model in
favor of a dynamical or variable dark energy EoS
$w_{DE}=P_{DE}/\rho_{DE}=w_{DE}(z)$ which evolves from a phantom
to a quintessence EoS during the late-time epoch
\cite{Cai:2025mas,Ye:2024ywg,Chaudhary:2025vzy,Chaudhary:2025uzr,Giare:2024smz,
Pan:2025qwy,Yang:2025mws,Zhang:2025bmk,Ong:2025utx,Nojiri:2025uew,OdintsovSGS_game:2024,OdintsovOS:2025,OdintsovOS:2026}.

Effective dynamical dark energy may be generated not only in
models with a given variable EoS, but also in modified gravity
theories, in particular, in $F(R)$ gravity theories with
non-trivial dependence on the Ricci scalar $R$ in the Lagrangian
\cite{Nojiri:2025uew,OdintsovSGS_game:2024,OdintsovOS:2025,OdintsovOS:2026,
Nojiri:2003ft,Capozziello:2005ku,Hwang:2001pu,Song:2006ej,Faulkner:2006ub,Olmo:2006eh,
Sawicki:2007tf,Faraoni:2007yn,Carloni:2007yv,Nojiri:2007as,Deruelle:2007pt,Appleby:2008tv,Linder:2009,
Dunsby:2010wg,Hu:2007nk,Bamba:2012qi,Odintsov:infl2023,OnofrioOS:2025,
OdintsovSGS:2017,OdintsovSGSlog:2019,OdintsovSGS_Axi:2023}. These
models are in general motivated by UV-completions of Einstein
gravity, and can unify the late-time epoch with the early-time
inflationary era.

Another approach in modelling modified gravity, which is also
motivated from the quantum effective theory, is non-local theory
of gravity
\cite{nonlocal,Deser:2007jk,Deffayet:2009ca,Deser:2013uya,Deser:2019lmm,NojiriO:2008,NojiriOSJ:2011,Joukovskaya:2007,Calcagni:2008,JhinganNOSTZ:2008,CapozzielloENO:2008,Nojiri:2019dio,NojiriOO:2026hij}.
In particular, it was shown in Ref.~\cite{Nojiri:2019dio} that a
non-local $F(R)$ gravity can be transformed into a ghost-free
local $F(R,\phi)$ model  by introducing a scalar field $\phi$. In
paper \cite{NojiriOO:2026hij} the inflationary stage of this model
was investigated and  predicted  inflationary parameters appeared
to be compatible with he latest Atacama Cosmology Telescope and
Planck constraints
\cite{AtacamaCosmologyTelescope:2025nti,Planck:2018jri,BICEP:2021}.

In this paper, we test the late-time dynamics of the $F(R,\phi)$ model
\cite{Nojiri:2019dio,NojiriOO:2026hij} originated from the non-local $F(R)$ gravity. We
confront this model with  Pantheon+ SNe Ia \cite{PantheonP:2022}, BAO DESI RD2
\cite{DESI:2025zgx}, the Hubble parameter observational data and compare this model with
the $\Lambda$CDM and the generalized exponential $F(R)$ scenarios.

This article is organized as follows: in section \ref{Dynamics},
the dynamical equations for $F(R,\phi)$ gravity during its
late-time evolution are described. In section \ref{Expon} the
generalized exponential $F(R)$ model is investigated. In the next
section the results of observational tests with SNe Ia, $H(z)$ and
BAO DESI data are described for both the considered models in
comparison with the $\Lambda$CDM scenario. In section \ref{CMB}
the models are confronted with the mentioned observations along
with CMB data. In the final section, the main results and
conclusions are presented.

\section{Dynamics of Ghost-free non-local Gravity and its Formulation in Terms of $F(R,\phi)$ Gravity}
\label{Dynamics}

In the article \cite{Nojiri:2019dio}, the non-local gravity model
with the action,
\begin{equation}
 \label{nlFRa1}
S= \int d^4 x \sqrt{-g} \left\{ \frac{1}{2\kappa^2} \left( R - \frac{1}{2} F(R)
\Box^{-1} F(R) \right) + \mathcal{L}_\mathrm{matter}  \right\} \, .
 \end{equation}
has been proposed. The model is based on $F(R)$ gravity, where $R$ is the Ricci scalar,
$\mathcal{L}_\mathrm{matter}$ is the matter Lagrangian density. This scenario has no
ghost degrees of freedom.

The gravitational  action \eqref{nlFRa1} with the  non-local
operator $\Box^{-1}$ in Ref.~\cite{NojiriOO:2026hij} was
transformed by using a scalar field $\phi$ to the following form,
\begin{equation}
 S= \int d^4 x \sqrt{-g} \left\{\frac{R}{2\kappa^2} -
\frac12\partial_\mu \phi \partial^\mu \phi - \phi F(R) f(R,\phi) +
\mathcal{L}_\mathrm{matter} \right\} \,.
 \label{actionmatter}\end{equation}
Following Ref.~\cite{NojiriOO:2026hij}, we use the notation,
\begin{equation}
\label{ffunction} f(R,\phi)=\frac{R}{\kappa^2} - \partial_\mu \phi \partial^\mu \phi -2
\phi F(R)
\end{equation}
with it derivatives $f_R$ and $f_{\phi}$:
 $$
f_R= \frac{\partial f(R,\phi)}{\partial R}=\frac{1}{\kappa^2} -2 \phi F'(R)\, , \qquad
f_{\phi}= \frac{\partial f}{\partial \phi}=-2 F(R)\, .
 $$
In the flat Friedmann-Lema\^{i}tre-Robertson-Walker (FLRW)
background metric,
\begin{equation}
\label{metricflrw} ds^2 = - dt^2 + a(t)^2 \sum_{i=1,2,3} \left(dx^i\right)^2\, ,
\end{equation}
the dynamical equations resulting from variations of the action
\eqref{actionmatter} with respect to the metric and the scalar
field $\phi$, take the form \cite{OdintsovO:2019},
 \begin{align}
\label{friedmaneq1}
3H^2 f_R =& \;\rho+\dot{\phi}^2 + \frac{Rf_R-f}{2}-3H\dot{f}_R \, , \\
\label{friedmaneq2}
 -(3H^2+2\dot{H})\,{f_R}=& \;P -\frac{Rf_R-f}{2}+\ddot{f}_R+2H\dot{f}_R\, , \\
\label{friedmaneq3} 0 = &\; \ddot{\phi}+3H\dot{\phi}+ F(R) \, .
\end{align}
Here, the ``dot'' denotes the derivative with respect to the
cosmic time $t$,  $\rho$ and $P$ are energy density and pressure
for dark matter and dark matter and radiation.

In the Ref. \cite{NojiriOO:2026hij}, the inflationary cosmology of
the model \eqref{actionmatter} was considered for the power-law
$F(R)$ function
 \begin{equation} \label{FRpower}
 F(R)=-\alpha R^n
 \end{equation}
for two cases, where the constant $n$ lies in the interval
$1<n<2$, and for $n=2$. Under the  slow-roll conditions  for the
Hubble rate $\dot{H}\ll H^2$ and for the scalar field
$\ddot{\phi}\ll H\dot{\phi}$ in  Ref.~\cite{NojiriOO:2026hij} the
slow-roll indices and the inflationary parameters have been
calculated and was shown that the spectral index of scalar
perturbation $n_s$ and the tensor-to-scalar ratio $r$ predicted in
the model \eqref{actionmatter}, \eqref{FRpower} satisfy the latest
Planck/BICEP and Atacama Cosmology Telescope (ACT) constraints
\cite{AtacamaCosmologyTelescope:2025nti,Planck:2018jri,BICEP:2021}.

In this article, we study the late-time evolution of the
considered $F(R,\phi)$ model that can be confronted with available
observational data. Unlike the inflationary epoch, at late times
the matter terms $\rho$ and $P$ in Eqs.~(\ref{friedmaneq1}),
(\ref{friedmaneq2}) are essential in the late-time dynamics of the
Universe. These equations lead to their standard FLRW evolution,
\begin{equation}
 \label{rhomr}
 \rho=\rho_m^{0}a^{-3}+\rho_r^{0}a^{-4}=\rho_m^{0}(a^{-3}+ X_r a^{-4})\, ,
\end{equation}
where we fix the present day radiation to matter ratio
\cite{OdintsovSGS_game:2024,OdintsovOS:2025}
$$X_r={\rho_r^{0}}\big/{\rho_m^{0}}=2.9656\cdot10^{-4}. $$
Equation \eqref{rhomr}, the relation,
\begin{equation}
 \label{RH}
R=12H^2 + 6\dot H
\end{equation}
together with the Eq.~(\ref{friedmaneq1}) in the form  \cite{NojiriOO:2026hij}
\begin{equation}
\label{friedmaneq_BB}
 \frac{3}{\kappa^2} H^2= \rho+
 \frac{\dot{\phi}^2}2 +\phi\left[F(R)- R F'(R)\right]  + 6 H \left[(H\phi+\dot\phi)\,F'(R) +\phi F''(R)\,\dot R \right]
\end{equation}
and  Eq.~(\ref{friedmaneq3}) form the system of dynamical equations for the considered
$F(R,\phi)$ model.

In these equations with the power-law $F(R)$ function \eqref{FRpower} we use the
dimensionless (normalized) variables $E$, ${\cal R}$, $\Phi$, $\Psi$
\begin{equation}
 \label{ERPhi}
E=\frac{H}{H_0},\qquad{\cal R}=\frac{R}{2\Lambda}, \qquad\Phi=\kappa\phi,
\qquad\Psi=\frac{\kappa\dot\phi}{H_0}
\end{equation}
instead of the Hubble parameter $H$, the Ricci scalar $R$, the functions $\phi$ and
$\dot\phi$. We also use the dimensionless free model parameters
\begin{equation}
 \label{OmmA}
\Omega_m^0=\frac{\kappa^2\rho_m^0}{3H_0^2}, \qquad A=2\kappa(2\Lambda)^{n-1}\alpha\,,
\end{equation}
representing the cold matter nowadays fraction and the constant
$\alpha$ in Eq.~(\ref{FRpower}) respectively. Here $H_0=H(t_0)$ is
the Hubble constant, the constant $\Lambda$ is associated with the
cosmological constant, though dark energy in this model is
generated by the $F(R)$ term and the scalar $\phi$, hence
$\Lambda$ is not considered as a free  parameter, but it is
expressed via $\Omega_m^0$ and $H_0$ as follows:
$$ \Omega_\Lambda=\frac{\Lambda}{3H_0^2}=1-\Omega_m^0(1+X_r)\,.
$$
In the notation of Eqs. (\ref{ERPhi}) and (\ref{OmmA}) with the
parameter $x=\log a$ instead of $t$ (where $\frac d{dt}=H\frac
d{dx}$) the dynamical equations (\ref{RH}), (\ref{friedmaneq_BB})
and (\ref{friedmaneq3}) can be rewritten in the following form,
\begin{eqnarray}
\frac{dE}{dx}&=&\Omega_\Lambda\frac{{\cal R}}{E}-2E\,,\label{eqH1} \\   
\frac{d{\cal R}}{dx}&=&\frac{\Omega_m^0(a^{-3}+ X_r a^{-4})-E^2+ \frac16\Psi^2 +A{\cal
R}^{n}\big[(n-1)\,\Omega_\Lambda\Phi-nE(E\Phi+\Psi)/E\big]}
{n(n-1)A\Phi E^2{\cal R}^{n-2}}\,,\\
\frac{d\Phi}{dx}&=&\frac1E\Psi\,,    \label{eqPhi1}\\
\frac{d\Psi}{dx}&=&-3\Psi+3A\Omega_\Lambda\frac{{\cal R}^n}E\,.     \label{eqPsi1}
  \end{eqnarray}
This system of equations can be integrated numerically over the
variable $x=\log a$ ``into the past'', if we start from the
initial conditions at the present time $t=t_0$ or $x=0$,
\begin{equation}
 \label{ERPhiinit}
E\big|_{x=0}=1,\qquad{\cal R}\big|_{x=0}={\cal R}_0, \qquad\Phi\big|_{x=0}=\Phi_0,
\qquad\Psi\big|_{x=0}=\Psi_0\,.
\end{equation}
These numerical solutions are determined if we fix seven free
model parameters including five constants (\ref{OmmA}),
(\ref{ERPhiinit}), $n$ and the Hubble constant,
\begin{equation} \label{7param}
\Omega_m^0,\quad n,\quad A,\quad{\cal R}_0,\quad\Phi_0,\quad\Psi_0,\quad H_0\,.
 \end{equation}
For any solution $E=E(a)$ and the fixed value $H_0$ we obtain the
Hubble parameter $H(a)=H_0E$ as a function of the scale factor or
$H=H(z)$ as a function of redshift
 $$z=\frac1a - 1$$
that can be compared with observational data.

For the considered $F(R,\phi)$ model, (\ref{actionmatter}),
(\ref{FRpower}) integral curves of the system
(\ref{eqH1})\,--\,(\ref{eqPsi1}) do not diverge into the ``past''
direction, this behavior differs from $F(R)$ models explored in
the papers
\cite{OdintsovSGS_game:2024,OdintsovOS:2025,OdintsovOS:2026} and
the generalized exponential $F(R)$ model (see Sect.~\ref{Expon}),
where we could integrate similar systems only ``into the future''.
This feature is the essential advantage of the $F(R,\phi)$ model.
However, viable cosmological solutions of the system
(\ref{eqH1})\,--\,(\ref{eqPsi1}) exist not for all values of model
parameters (\ref{7param}). For example, to exclude singularities
in the past we should choose only negative values of $\Phi_0$
(with large $|\Phi_0|$) in the initial conditions
(\ref{ERPhiinit}) and fix other model parameters under some
restrictions. These restrictions become more narrow if we confront
model predictions with observational data, including the Pantheon+
catalogue of Type Ia supernovae (SNe Ia) \cite{PantheonP:2022},
the DESI measurements of baryon acoustic oscillations
\cite{DESI:2025zgx} and the Hubble parameter estimations $H(z)$.
  \begin{figure}[th]
   \centerline{ \includegraphics[scale=0.68,trim=5mm 0mm 2mm -1mm]{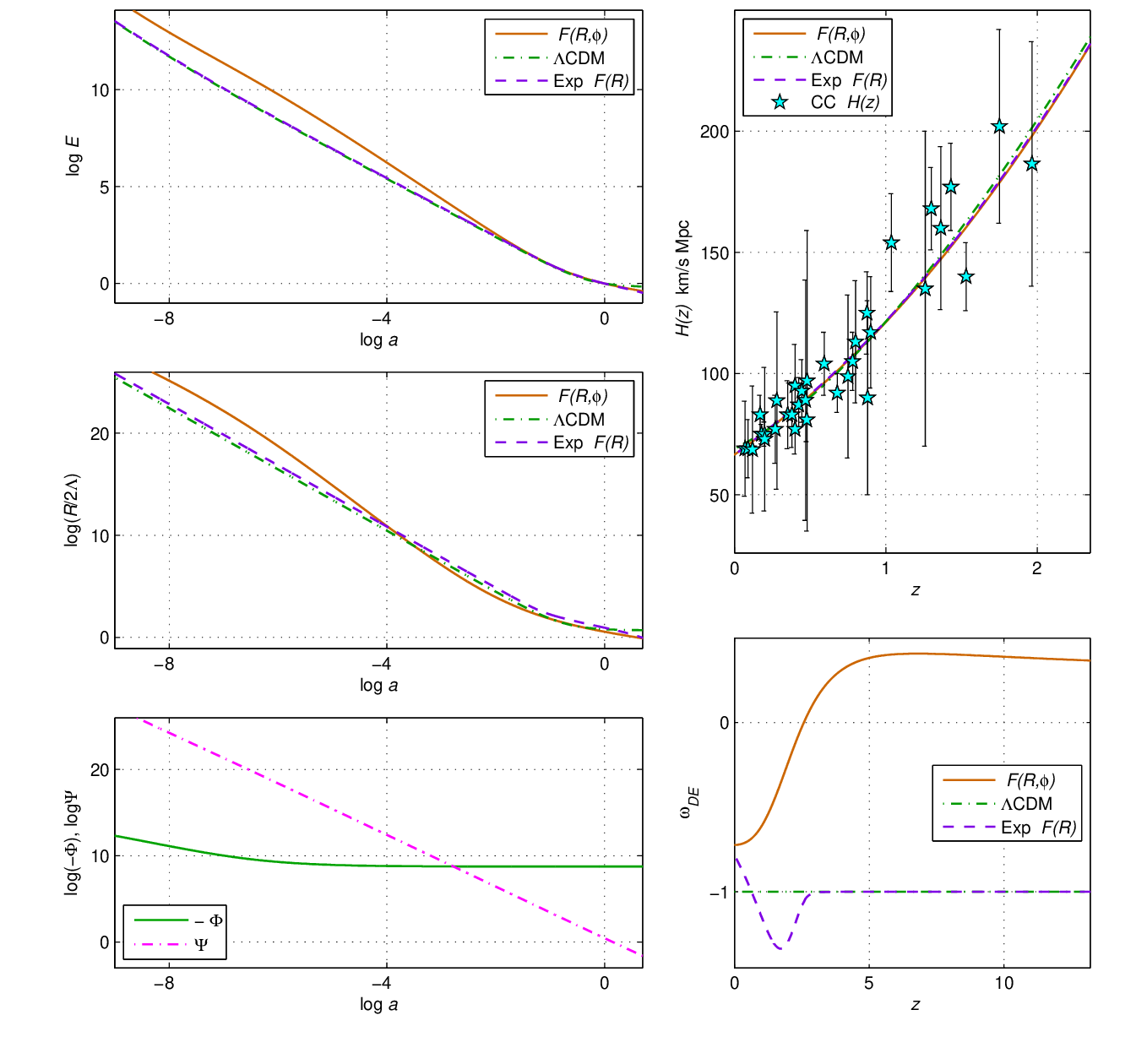}}
\caption{Evolution of the normalized  Hubble parameter $E$, Ricci scalar ${\cal R}$ and
the scalar field $-\Phi$,  $\Psi$ in logarithmic scale for the $F(R,\phi)$ model
(\ref{actionmatter}), (\ref{FRpower}) (the left panels), the  Hubble parameter $H(z)$
and the dark energy EoS parameter $\omega_{DE}(z)$ as functions of redshift (the right
panels)  in comparison to the $\Lambda$CDM model (\ref{LCDM}) and the generalized
exponential $F(R)$ model (\ref{GenExp}). The model parameters for 3 models are fixed
from Table~\ref{Est1}. }
  \label{F1}
\end{figure}
The calculated solutions $H(z)$ are tested with these observations
and we obtain the best fit values of the free model parameters
(\ref{7param}) for the $F(R,\phi)$ model. These best fits  are
tabulated in Table~\ref{Est1} below where we compare the results
with two other models.

The behavior of these solutions is shown in Fig.~\ref{F1} where
the evolution of the normalized Hubble parameter $E=H/H_0$, Ricci
scalar ${\cal R}=\frac{R}{2\Lambda}$ and the scalar field
parameters $\Phi$, $\Psi$ is presented in comparison with similar
variables in the $\Lambda$CDM model,
\begin{equation}\label{LCDM}
 H^2 =H_0^2\big[\Omega_m^0(a^{-3}+ X_r a^{-4})+\Omega_\Lambda\big]
\end{equation}
and the generalized exponential $F(R)$ model described below in
section \ref{Expon}. For all the models, the best fit parameters
from Table~\ref{Est1} are used in Fig.~\ref{F1}. In the left
panels we draw logarithms: $\log E$, $\log {\cal R}$,
$\log(-\Phi)$ and $\log \Psi$ as functions of $x=\log a$. One can
see that for for the $F(R,\phi)$ model the Ricci scalar ${\cal R}$
at $z>100$ ($a<10^{-2}$) exceeds this value for the $\Lambda$CDM
model. The similar behavior takes place for $E(a)$, but for the
best fit solution in the $F(R,\phi)$ model this difference is
reduced to minimal values.

In the top-right panel of Fig.~\ref{F1} the Hubble parameter
$H(z)$ depending on redshift $z$ is depicted for three models and
is compared with $H(z)$ observational data. And in the
bottom-right panel the dark energy EoS parameter,
\begin{equation}\label{EoSDE}
  \omega_{DE}(z)=\frac{P_{DE}}{\rho_{DE}}=-1+\frac13\frac{d}{dx}\log\big[E^2-\Omega_m^0(a^{-3}+ X_r a^{-4})\big]\,.
\end{equation}
is shown for the same best fitted $E(z)$. Here $\rho_{DE}$ is the
effective DE density generated by $\phi$ and $F(R)$ gravity,
$\rho_{DE}$ is the additional summand to $\rho$ in the right hand
side of Eq.~(\ref{friedmaneq_BB}). The effective DE pressure
$P_{DE}$ cam similarly be extracted from Eq.~(\ref{friedmaneq2}).
For the $\Lambda$CDM model the dark energy EoS parameter is
constant $\omega_{DE}=-1$, but for the model (\ref{actionmatter}),
(\ref{FRpower}) it behaves as variable quintessential EoS
$\omega_{DE}>-1$ diminishing at late time ($z<5$) to
$\omega_{DE}\big|_{z=0}\simeq-0.724$. This behavior differs from
variable EoS in other $F(R)$ models
\cite{OdintsovSGS_game:2024,OdintsovOS:2025,OdintsovOS:2026},
where $\omega_{DE}(z)$ for the best fitted solutions evolute from
a phantom to a quintessential stage at $z<2$.

Note that the normalized Hubble parameter $E(x)$ for the
$F(R,\phi)$ model (\ref{actionmatter}), (\ref{FRpower}) in the
top-left panel of Fig.~\ref{F1} exceeds the $\Lambda$CDM $E(x)$ at
high redshifts or $x\to-\infty$. This difference is an inevitable
feature of the $F(R,\phi)$ model, and the best description of the
observations is achieved when this difference is minimal. The
evolution of the scalar field rate parameter (\ref{ERPhi})
$\Psi={\kappa\dot\phi}/{H_0}$ looking like the straight line in
the bottom-left panel of Fig.~\ref{F1} shows that it behaves
approximately as $\Psi\simeq\Psi_0a^{-3}$. In other words, the
last term $3A\Omega_\Lambda{\cal R}^n/E$ in the right hand side of
Eq. (\ref{eqPsi1}) appeared to be vanishing for the best fit
parameters.

\section{Generalized exponential $F(R)$ model}
 \label{Expon}

The generalized exponential $F(R)$ model with the Lagrangian,
\begin{equation}\label{GenExp}
    F(R)=R+\frac{R^2}{M^2}-\Lambda  \left[2 -
    \alpha\exp\left(-\beta\frac{R}{2\Lambda}\right)\right]\,    .
\end{equation}
was considered in Ref.~\cite{OdintsovOS:2025}. Here $\alpha$ and $\beta$ are positive
constants, $F_\mathrm{inf}(R)=\frac{R^2}{M^2}$ is the inflationary term, it is assumed
to be negligible near and after the recombination epoch. This model has the
$\Lambda$CDM-like asymptotic behavior at the large $R$ limit, in other words, its
Lagrangian tends to the $\Lambda$CDM expression $F(R)\approx R -2\Lambda$ at the epoch,
when $R\gg\Lambda$, bur far later the inflationary era, where $F_\mathrm{inf}(R)$
remains negligible.

The dynamics of the model (\ref{GenExp}) and other scenarios with
similar asymptotic behavior was described in Refs.
\cite{OdintsovOS:2025,OdintsovOS:2026} and it reduces to the
relation \eqref{RH}  $R=6\dot H + 12H^2$ and the Friedmann
equation
 \begin{equation}
\frac{dR}{d\log
a}=\frac1{F''(R)}\bigg(\frac{\kappa^2\rho}{3H^2}-F'(R)+\frac{RF'(R)-F}{6H^2}\bigg)\ .
   \label{eqR1}
  \end{equation}
These equations in the notation  of Eq. \eqref{ERPhi} may be
reduced to the system including Eq.~\eqref{eqH1} and the equation,
\begin{equation}
\frac{d{\cal R}}{dx}=2\frac{\big[\Omega_m^{0}(a^{-3}+ X_r a^{-4})
+\Omega_\Lambda\big(1-\frac12\alpha(1+\beta{\cal R})\,e^{-\beta{\cal
R}}\big)\big]\big/E^2-1+\frac12\alpha\beta e^{-\beta{\cal R}}}
 {\alpha\beta^2 e^{-\beta{\cal R}}}\,.
  \label{eqR2}
  \end{equation}
This system should be integrated numerically, but unlike the
$F(R,\phi)$ model (\ref{actionmatter}), in this scenario only the
future direction for integrating is acceptable (with growing $a$
or $x$), because in the opposite direction the integral curves of
the system (\ref{eqH1}), (\ref{eqR2}) diverge and deviate from
viable solutions. Hence, we can not start from the present time
and have to define initial conditions for the system (\ref{eqH1}),
(\ref{eqR2}) at some point $a_\mathrm{ini}$ or equivalently
$z_\mathrm{ini}=a_\mathrm{ini}^{-1}-1$ in the past. This initial
point for the considered model \eqref{GenExp} is determined from
the condition that the term $F''(R)$ in the denominator of
Eq.\~(\ref{eqR1}) should be very small, but not negligible. More
precisely, the dimensionless factor
$\delta=\alpha\beta^2e^{-\beta{\cal R}_\mathrm{ini}}$ in the
denominator of Eq.\~(\ref{eqR2}) should be  much smaller than
unity.

At the initial point $a_\mathrm{ini}$ and before, the solutions
$H(a)$, ${\cal R}(a)$ of the $F(R)$ model (\ref{GenExp}) should
have the $\Lambda$CDM-like asymptotic behavior \eqref{LCDM}
\cite{OdintsovSGS_game:2024,OdintsovOS:2025,OdintsovOS:2026}:
 \begin{equation} \label{asymLCDM}
 \frac{H^2}{H^{2}_0}=\Omega_m^{0} \big(a^{-3}+ X_r a^{-4}\big)+\Omega_\Lambda\,,\qquad
 {\cal R}=\frac{R}{2\Lambda}=2+\frac{\Omega_m^{0}}{2\Omega_\Lambda}a^{-3}\ .
 \end{equation}
To determine $a_\mathrm{ini}$ we assume  $\delta\sim 10^{-9}$ and
obtain,
\begin{equation}\label{aini}
a_\mathrm{ini}=\bigg[\frac{2\Omega_\Lambda}{\Omega_m^0}\bigg(\frac{\log(\alpha\beta^2/\delta)}{\beta}-2\bigg)\bigg]^{-1/3}
.
\end{equation}
During integration of the system (\ref{eqH1}), (\ref{eqR2}),  we
should also solve another problem: at the starting point
$a_\mathrm{ini}$ we do not know the true value of the Hubble
constant $H_0$ and therefore the parameters $\Omega_m^{0}$,
$\Omega_\Lambda$ are unknown. Following Refs.
\cite{OdintsovSGS_game:2024,OdintsovOS:2025,OdintsovOS:2026} we
introduce a ``preliminary''  $\Lambda$CDM-asymptotical Hubble
constant $H^*_0$  at the initial point, that differs from the true
Hubble constant $H_0=H(t_0)$ achieved during evolution in this
scenario from $a_\mathrm{ini}$ to the present day value $a=1$. The
value $H^*_0$ determines the normalized Hubble rate,
 \begin{equation}
 \label{Enorm}
E^*=\frac{H}{H_0^*}\,,
\end{equation}
 and also the parameters,
\begin{equation} \label{Omega_mL2}
\Omega_m^{*}=\frac{\kappa^2\rho_m^0}{(H_0^*)^2}\,,\qquad
\Omega_\Lambda^*=\frac{\Lambda}{3(H_0^*)^2}\,.
 \end{equation}
When we integrate  the system with the modified equation
(\ref{eqH1}),
 \begin{equation}
\frac{dE^*}{dx}=\Omega_\Lambda^*\frac{{\cal R}}{E^*}-2E^*\,,\label{eqH2}
 \end{equation}
and Eq. (\ref{eqR2}), with $\Omega_m^0/E^2=\Omega_m^{*}/(E^*)^2$
and $\Omega_\Lambda/E^2=\Omega_\Lambda^{*}/(E^*)^2$, we obtain the
solution $E^*(a)$, leading to the Hubble parameter
$H(a)=H_0^*E^*(a)$. Thus we reconstruct the Hubble constant
$H_0=H|_{a=1}=H_0^*E^*|_{a=1}$ and the parameters $\Omega_m^{0}$,
$\Omega_\Lambda$ from the relations,
 \begin{equation} \label{H0Omm}
 \Omega_m^0H_0^2=\Omega_m^{*}(H^{*}_0)^2=\kappa^2\rho_m^0\,,
 \qquad  \Omega_\Lambda H_0^2=\Omega_\Lambda^{*}(H^{*}_0)^2=\frac{\Lambda}3\,,
 \qquad \frac{\Omega_\Lambda}{\Omega_m^0}=\frac{\Omega_\Lambda^*}{\Omega_m^*}\;.
 \end{equation}
These solutions are shown in Fig.~\ref{F1} and are tested with the
observational data.

\section{Observational data tests}
 \label{Tests}

We now describe the process of using and interpreting the
observational data, by using the line of research developed in
previous studies
\cite{OdintsovSGS_game:2024,OdintsovOS:2025,OdintsovOS:2026}. For
the Pantheon+ catalog \cite{PantheonP:2022} of Type Ia supernovae
(SNe Ia) data  with with $N_{\mathrm{SN}}=1701$ datapoints of the
distance moduli $\mu_i^\mathrm{obs}$ at redshifts $z_i$ we
calculate the $\chi^2$ function:
 \begin{equation}
\chi^2_{\mathrm{SN}}(\theta_1,\dots)=\min\limits_{H_0} \sum_{i,j=1}^{N_\mathrm{SN}}
 \Delta\mu_i\big(C_{\mathrm{SN}}^{-1}\big)_{ij} \Delta\mu_j\ ,\qquad 
 \Delta\mu_i=\mu^\mathrm{th}(z_i,\theta_1,\dots)-\mu^\mathrm{obs}_i\ .
 \label{chiSN}
\end{equation}
Here $C_{\mbox{\scriptsize SN}}$ is the covariance matrix
\cite{PantheonP:2022} and the theoretical estimates of the
distance moduli are made as follows,
\begin{equation}
 \mu^\mathrm{th}(z) = 5 \log_{10} \frac{(1+z)\,D_M(z)}{10\mbox{pc}},\qquad D_M(z)= c \int\limits_0^z\frac{d\tilde z}{H(\tilde
 z)}.    \label{muDM}
\end{equation}
For baryon acoustic oscillations data from the DESI 2025
\cite{DESI:2025zgx}, we calculate the values,
 $$
\frac{D_M(z)}{r_d},\qquad\frac{D_H(z)}{r_d}=\frac{c}{H(z)\,r_d},\qquad
\frac{D_V(z)}{r_d}=\frac{(zD_H D_M^2)^{1/3}}{r_d},
 $$
where $r_d=r_s(z_d)$  is THE comoving sound horizon at the end of
the baryon drag era, calculated in accordance with Refs.~
\cite{OdintsovSGS_game:2024,OdintsovOS:2025,OdintsovOS:2026} as
the integral,
 \begin{equation}
r_s(z)=  \int_z^{\infty} \frac{c_s(\tilde z)}{H (\tilde z)}\,d\tilde
z=\frac1{\sqrt{3}}\int_0^{1/(1+z)}\frac{da}
 {a^2H(a)\sqrt{1+\big[3\Omega_b^0/(4\Omega_\gamma^0)\big]a}}\ ,
  \label{rs2}\end{equation}
We use BAO DESI data \cite{DESI:2025zgx} with the observed value
$D_V(z_1)/r_d$ at $z_1=0.295$ and data points with $D_M(z_i)/r_d$
and $D_H(z_i)/r_d$ for higher redshifts $z_i$ and we calculate the
$\chi^2$ function,
\begin{equation}
\chi^2_{\mathrm{BAO}}(\theta_1,\dots)=\bigg[\frac{\Delta_V(z_1)}{\sigma_V(z_1)}\bigg]^2
+\sum_{i=2}^8 [\Delta_M(z_i)\;\,\Delta_H(z_i)]\,C^i_{M,H}
\bigg[\begin{array}{c}\!\Delta_M(z_i)\!\\ \Delta_H(z_i)\end{array}\bigg],
 \label{chiBAO}\end{equation}
where,
$\Delta_q=\big(\frac{D_q}{r_d}\big)^\mathrm{th}-\big(\frac{D_q}{r_d}\big)^\mathrm{obs}$,
 $q=V,\,M,\,H$; $C^i_{M,H}$ are the covariance matrices with the errors
$\sigma_q(z_1)$ and the cross-correlation coefficients  $r^i_{M,H}$ between
$D_M(z_i)/r_d$ and $D_H(z_i)/r_d$.

\begin{figure}[th]
   \centerline{ \includegraphics[scale=0.68,trim=5mm 0mm 2mm -1mm]{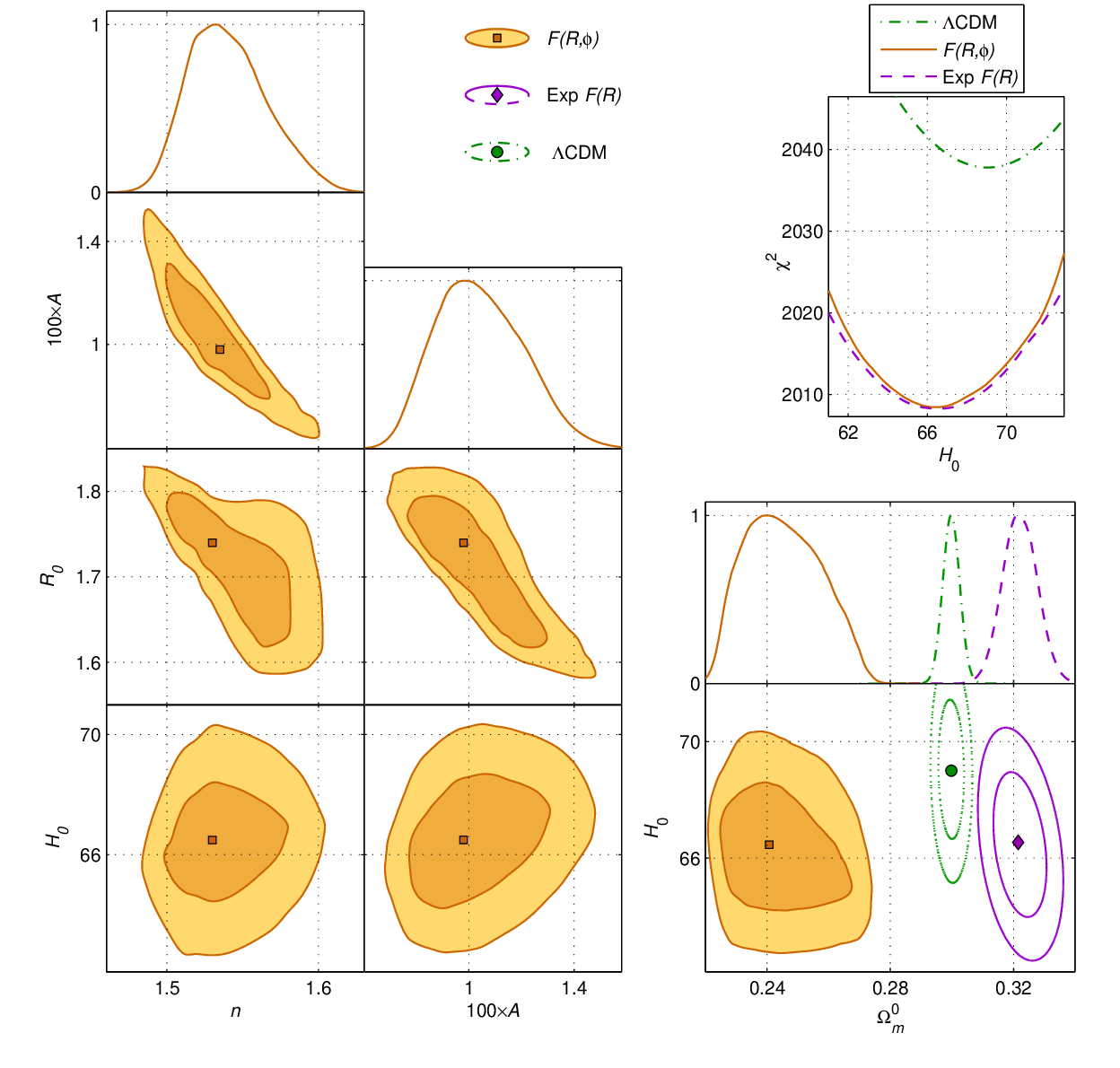}}
\caption{Contour plots of $\chi^2$ with $1\sigma$, $2\sigma$ CL,
likelihood functions $ {\cal L}(\theta_i)$ and one-parameter
distributions $\chi^2(H_0)$ for the $F(R,\phi)$ model
(\ref{actionmatter}), (\ref{FRpower}) in comparison with the
$\Lambda$CDM model (\ref{LCDM}) and the $F(R)$ model
\eqref{GenExp} for SNe Ia, BAO DESI and $H(z)$ data. }
  \label{F2}
\end{figure}

For the Hubble parameter data $H(z)$ we use here $N_H=34$
datapoints of $H^\mathrm{obs}(z_i)$ (Cosmic Chronometers)
tabulated in Ref. \cite{Pan:2025qwy} and we calculate the
corresponding $\chi^2$ function,
 \begin{equation}\label{chiH}
\chi^2_{H}= \sum_{i=1}^{N_H} \left[\frac{H^\mathrm{obs}(z_i)
 -H^\mathrm{th}(z_i; \theta_k)}{\sigma_{H,i}}\right]^2 \, .
 \end{equation}
To determine the best fit model parameters of the considered
scenarios, we calculate and minimize the total $\chi^2$ function
with the contributions from SNe Ia (\ref{chiSN}), BAO DESI
(\ref{chiBAO}) and the Hubble parameter data (\ref{chiH}):
 \begin{equation}
  \chi^2=\chi^2_\mathrm{SN}+\chi^2_\mathrm{BAO}+\chi^2_H\ .
 \label{chitot} \end{equation}
The results of these calculations for the $F(R,\phi)$ model
(\ref{actionmatter}), (\ref{FRpower}) in comparison to the
$\Lambda$CDM scenario (\ref{LCDM}) and the generalized exponential
$F(R)$ model \eqref{GenExp} are presented in Table~\ref{Est1} and
in Fig.~\ref{F2}. One can see that the $F(R,\phi)$ model describes
this data more successfully than the $\Lambda$CDM, not only in the
achieved value of $\min\chi^2$ (2008.43 vs 2037.79), but also
after consideration of the Akaike information criterion (AIC)
 \cite{Liddle_ABIC:2007}
 \begin{equation}
 \mbox{AIC} = \min\chi^2 +2N_p\,.
  \label{AIC}\end{equation}
Here $N_p$ is the number of free model parameters: $N_p=7$ (see
(\ref{7param})) for the model (\ref{actionmatter}),
(\ref{FRpower}) and $N_p=2$ for the  $\Lambda$CDM scenario
(\ref{LCDM}). This criterion brings an additional advantage to
scenarios with small number $N_p$, but this advantage appeared to
be insufficient to change the resulting physical picture.
\begin{table}[ht]
\begin{center}
\caption{Best fits with $1\sigma$ errors, $\min\chi^2$, AIC,  from SNe Ia, BAO DESI and
$H(z)$ data, for the $F(R,\phi)$ model (\ref{actionmatter}), (\ref{FRpower}) in
comparison with the  $\Lambda$CDM model (\ref{LCDM}) and the generalized exponential
$F(R)$ model \eqref{GenExp}.}
\begin{tabular}{|l|c|c|c|c|c|c|}  \hline
 \hline  Model &   $\min\chi^2/d.o.f$& AIC & $\Omega_m^0$& $H_0$&   other parameters \\
\hline
 $F(R,\phi)$ & 2008.43 /1736 & 2022.43& $0.2406^{+0.0198}_{-0.0128}$ & $66.39^{+1.39}_{-1.40}$  &
  $\begin{array}{c} n=1.53_{-0.021}^{+0.033},\;\,A=0.0098_{-0.0014}^{+0.0024},\;\,{\cal R}_0=1.74_{-0.11}^{+0.06},\\
   \Psi_0=1.615_{-0.093}^{+0.045},\;\,\Phi_0=-6340_{-74}^{+68} \end{array} $ \rule{0pt}{1.9em}  \\
\hline
$\Lambda$CDM& 2037.79 /1741 & 2041.79& $0.2998^{+0.0027}_{-0.0028}$& $69.01^{+1.61}_{-1.54}$ & -  \rule{0pt}{1.1em}  \\
\hline
 Exp $F(R)$ & 2008.18 /1738 & 2018.18& $0.3213^{+0.0065}_{-0.0052}$ & $66.46^{+1.60}_{-1.60}$  & $\alpha=2.76_{-1.42}^{+3.34}$,
$\beta=0.694_{-0.237}^{+0.196}$, $\Omega_\Lambda=0.545_{-0.074}^{+0.062}$ \rule{0pt}{1.1em}  \\
\hline
  \hline \end{tabular}
 \end{center}
 \label{Est1}
\end{table}
Fig.~\ref{F2} illustrates our analysis of the $\chi^2$ function
(\ref{chitot}) for the $F(R,\phi)$ model with contour plots at
$1\sigma$,  $2\sigma$ confidence levels (CL) for two-parameter
distributions $\chi^2(\theta_j,\theta_k)$ with pairs of free
parameters (\ref{7param}).
For each pair of chosen model parameters to calculate
$\chi^2(\theta_j,\theta_k)$ we seek for the minimum of $\chi^2$
over the remaining five parameters. In this process, the grid
spacing and the size of the box for other parameters are
determined at the initial stage, but the center of the box is
approximated during this process. The prior ranges for the model
parameters have their physical limits, in particular, for the
$F(R,\phi)$ model (\ref{actionmatter}), (\ref{FRpower}) they are,
\begin{equation}\label{Priors1}
 \Omega_m^0\in[0.05,0.5];\;\; n\in(1,2];\;\;A\in[0,1];\;\;{\cal R}_0\in[0.5,5];\;\;
\Phi_0\in[-10^5,-50];\;\; \Psi_0\in[0,5];\;\;H_0
H_0\in[50,100]\;\,\frac{\mathrm{km}}{\mathrm{s\cdot Mpc}}\,.
\end{equation}
In the top-right panel in Fig.~\ref{F2} one-parameter
distributions are presented, with,
 $$ 
 \chi^2(H_0)=\min\limits_{\mathrm{other}\;\theta_j} \chi^2(\theta_1,\theta_2,\dots,H_0)\,.
 $$ 
for the aforementioned three models. One can see that the
$F(R,\phi)$ and $F(R)$ models (\ref{actionmatter}), \eqref{GenExp}
are much more successful regarding the minimum of $\chi^2$, which
is essentially lower than the $\Lambda$CDM model. The estimates of
$\min\chi^2$ in Table~\ref{Est1} show the difference
$\Delta\min\chi^2$ of order 30 in favor of the $F(R,\phi)$ and
$F(R)$ models. This advantage is kept also for the Akaike
information criterion: $\Delta\mathrm{AIC}\simeq-19.36$ for the
$F(R,\phi)$ and $\Delta\mathrm{AIC}\simeq-23.61$ for the $F(R)$
model.

The likelihood functions ${\cal L}(\theta_j)$ for the parameters
$\theta_j$ shown in Fig.~\ref{F2} are related with the
corresponding one-parameter distributions $\chi^2(\theta_j)$:
   \begin{equation}
{\cal L}(\theta_j)= \exp\bigg[- \frac{\chi^2(\theta_j)-m_\mathrm{abs}}2\bigg]\ ,
 \label{likeli} \end{equation}
where $m_\mathrm{abs}$ is the absolute minimum for $\chi^2$. The
squares for the $F(R,\phi)$ models, circles and diamonds for other
models, denote the best fits with $\min\chi^2$ of the
corresponding $\chi^2(\theta_j,\theta_k)$. We see that the best
fits for the exponent $n$ lie in the narrow range $1.5<n<1.6$, the
best fit for $A$ is of order 0.01, suitable interval for $\Phi_0$
is negative: $\Phi_0=-6340_{-74}^{+68}$.

In the bottom-right panel of Fig.~\ref{F2} with contours in the
$\Omega_m^0-H_0$ plane we compare the $F(R,\phi)$ model
(\ref{actionmatter}), (\ref{FRpower}) with two other models. One
may conclude that the best $\Omega_m^0$ fits for the $F(R,\phi)$
model are essentially lower compared to other scenarios. This
difference may be related with the different physical significance
of $\Omega_m^0$ and of dark energy in these scenarios.
 \begin{figure}[th]
   \centerline{ \includegraphics[scale=0.68,trim=5mm 0mm 2mm -1mm]{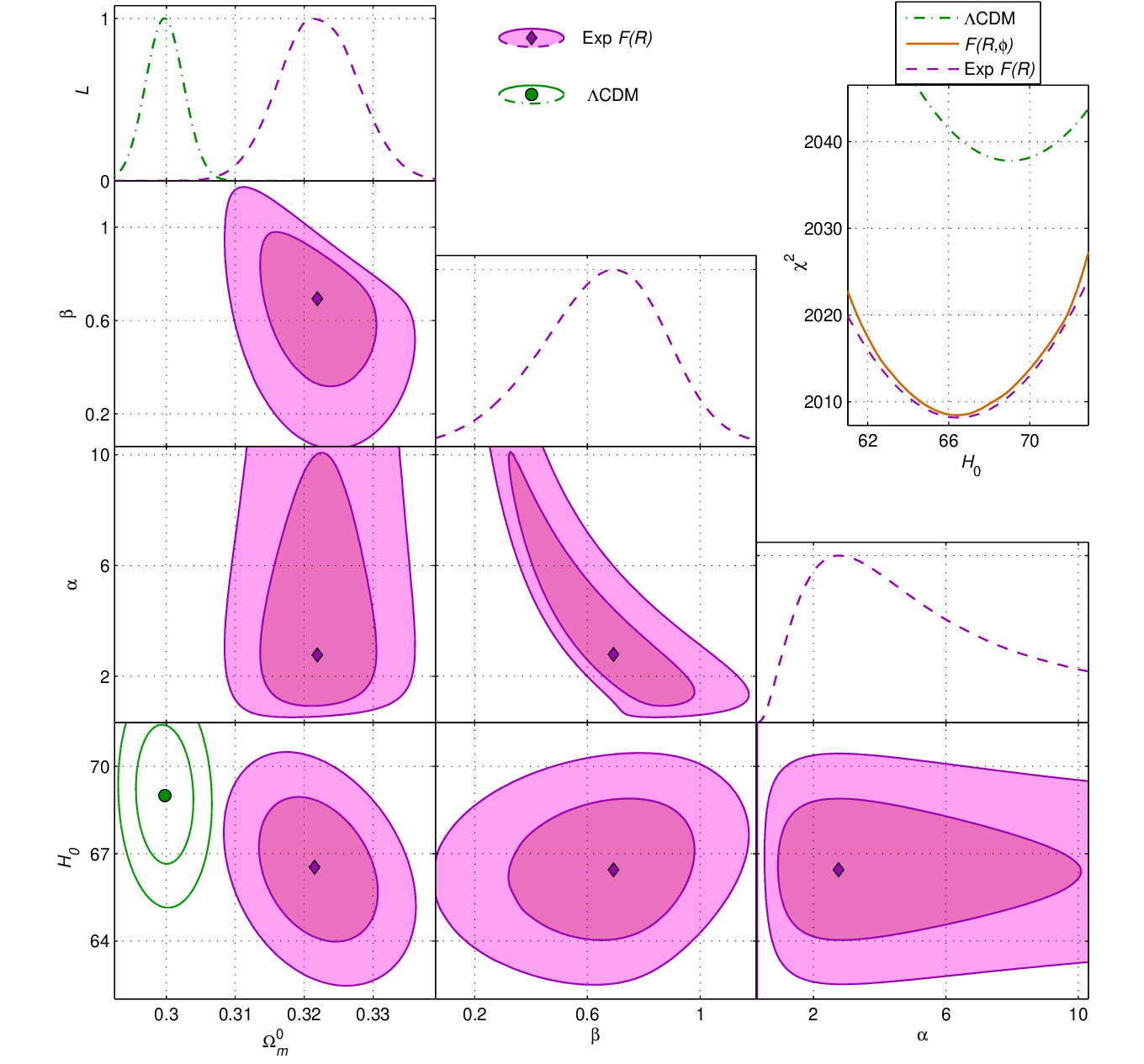}}
\caption{Contour plots of $\chi^2$ with $1\sigma$, $2\sigma$ CL, likelihood functions $
{\cal L}(\theta_i)$ and one-parameter distributions $\chi^2(H_0)$ for  the generalized
exponential $F(R)$ model \eqref{GenExp} in comparison with models (\ref{actionmatter})
and (\ref{LCDM})  for SNe Ia, BAO DESI and $H(z)$ data. }
  \label{F3}
\end{figure}
The similar analysis of observational tests with SNe Ia, BAO DESI
and $H(z)$ data for the generalized exponential $F(R)$ model
\eqref{GenExp} is illustrated in Fig.~\ref{F3}. This model is the
most successful in $\min\chi^2$ and AIC having $N_p=5$ parameters,
though the resulting $\chi^2$ in this case, is close to that of
the $F(R,\phi)$ model. The prior ranges for the scenario
\eqref{GenExp} coincide with \eqref{Priors1} for  $\Omega_m^0$ and
$H_0$ and $\alpha\in[0,30]$; $\beta\in[0,10]$;
$\Omega_\Lambda\in[0.4,1]$.

In the $\Omega_m^0-H_0$ plane, the difference between the best
fits for $\Omega_m^0$ exceeds $3\sigma$: it can explain the
advantage in $\min\chi^2$ over the $\Lambda$CDM model for the
chosen set of observational data. This success of the $F(R)$
scenario is also connected with behavior of its effective dark
energy EoS parameter $\omega_{DE}(z)$ \eqref{EoSDE} at redshift
$0<z<3$ shown in  Fig.~\ref{F1}. As one can see, $\omega_{DE}(z)$
evolves from a phantom to a quintessence stage. The difference in
the best fits for $H_0$ between the $F(R)$ and $\Lambda$CDM models
is not so striking, but also essential, since the considered
$F(R,\phi)$ and $F(R)$ models predict the Hubble constant $H_0$ of
the order $66.4$ km/s/Mpc. The $2\sigma$ CL domains for the
parameters $\alpha$ and $\beta$ are extended to larger $\alpha>10$
and small $\beta\sim0.05$. These features depend on chosen
observational datasets, and in the next section we will see how
this picture changes with the addition of the CMB data in our
statistical analysis.

\section{Observational Tests with the Addition of CMB Data}
 \label{CMB}

The considered $F(R,\phi)$ model (\ref{actionmatter}),
(\ref{FRpower}) and other scenarios should be also confronted with
observational data coming from the CMB. In this article, we use
the CMB observational parameters in accordance with
Refs.~\cite{OdintsovSGS_game:2024,OdintsovOS:2025,OdintsovOS:2026}
 $$
\mathbf{x}=\left(R,\ell_A,\omega_b \right)\, ,\quad
R=\sqrt{\Omega_m^0}\frac{H_0D_M(z_*)}c\, ,\quad \ell_A=\frac{\pi D_M(z_*)}{r_s(z_*)}\, ,
\quad\omega_b=\Omega_b^0h^2
  $$
 with the Planck 2018 data priors \cite{Planck:2018,ChenHW:2018}
 $$
\mathbf{x}^\mathrm{Pl}=\left( R^\mathrm{Pl},\ell_A^\mathrm{Pl},\omega_b^\mathrm{Pl}
\right) =\left( 1.7428\pm0.0053,\;301.406\pm0.090,\;0.02259\pm0.00017 \right)
$$
for scenarios with zero spatial curvature. Here, the comoving
sound horizon $r_s(z_*)$ is calculated as the integral \eqref{rs2}
where the redshift $z_*\simeq1090$ is related to the
photon-decoupling epoch. The value $z_*$ is estimated following
Refs.~\cite{OdintsovSGS_game:2024,OdintsovSGS_Axi:2023,ChenHW:2018}.
We calculate the $\chi^2$ function with the covariance matrix
$C_{\mathrm{CMB}}=\| \tilde C_{ij}\sigma_i\sigma_j \|$
\cite{ChenHW:2018}
 $$
\chi^2_\mathrm{CMB}=\min_{\omega_b,H_0}\Delta\mathbf{x}\cdot C_{\mathrm{CMB}}^{-1}\left(
\Delta\mathbf{x} \right)^{T}\, ,\quad \Delta
\mathbf{x}=\mathbf{x}-\mathbf{x}^\mathrm{Pl}\,.
 $$
and the total $\chi^2$ with the following four sources of
observational data,
 \begin{equation}
  \chi^2_\mathrm{tot}=\chi^2_\mathrm{SN}+\chi^2_\mathrm{BAO}+\chi^2_H+\chi^2_\mathrm{CMB}\ .
 \label{chi4} \end{equation}
Calculations of $\chi^2_\mathrm{tot}$ including the CMB data for
the $F(R,\phi)$ model (\ref{actionmatter}), (\ref{FRpower}) led to
rather disappointing results, since the values
$\chi^2_\mathrm{tot}$ appeared to be extremely large. These
results are directly related with the behavior of $F(R,\phi)$
solutions shown in  Fig.~\ref{F1}; if the chosen model parameters
\eqref{7param} satisfy the SNe Ia, BAO DESI and $H(z)$ CC
limitations, the resulting normalized Hubble rate $E(z)$ at
$z\sim1000$ (or $a\sim10^{-3}$) essentially exceeds $E(z)$ for the
$\Lambda$CDM and generalized exponential models. This behavior
allows us to describe BAO DESI data with some difficulty because
of the integral \eqref{rs2} for $r_d$, but it becomes an
overpowering obstacle after including the CMB data related to
$z\sim1100$.

From another side, we can describe the CMB data with the
$F(R,\phi)$ model (\ref{actionmatter}), (\ref{FRpower})
successfully if we choose the starting point $a_\mathrm{ini}$ for
the integration near $a\sim10^{-3}$. But in this case the SNe Ia,
BAO DESI and CC data related to $z\in[0,2.4]$ appear to be
described badly with large $\chi^2$ \eqref{chitot}. One may
conclude that the $F(R,\phi)$ model (\ref{actionmatter}),
(\ref{FRpower}) needs some corrections for successful tests with
all considered observational data.

\begin{figure}[th]
   \centerline{ \includegraphics[scale=0.68,trim=5mm 0mm 2mm -1mm]{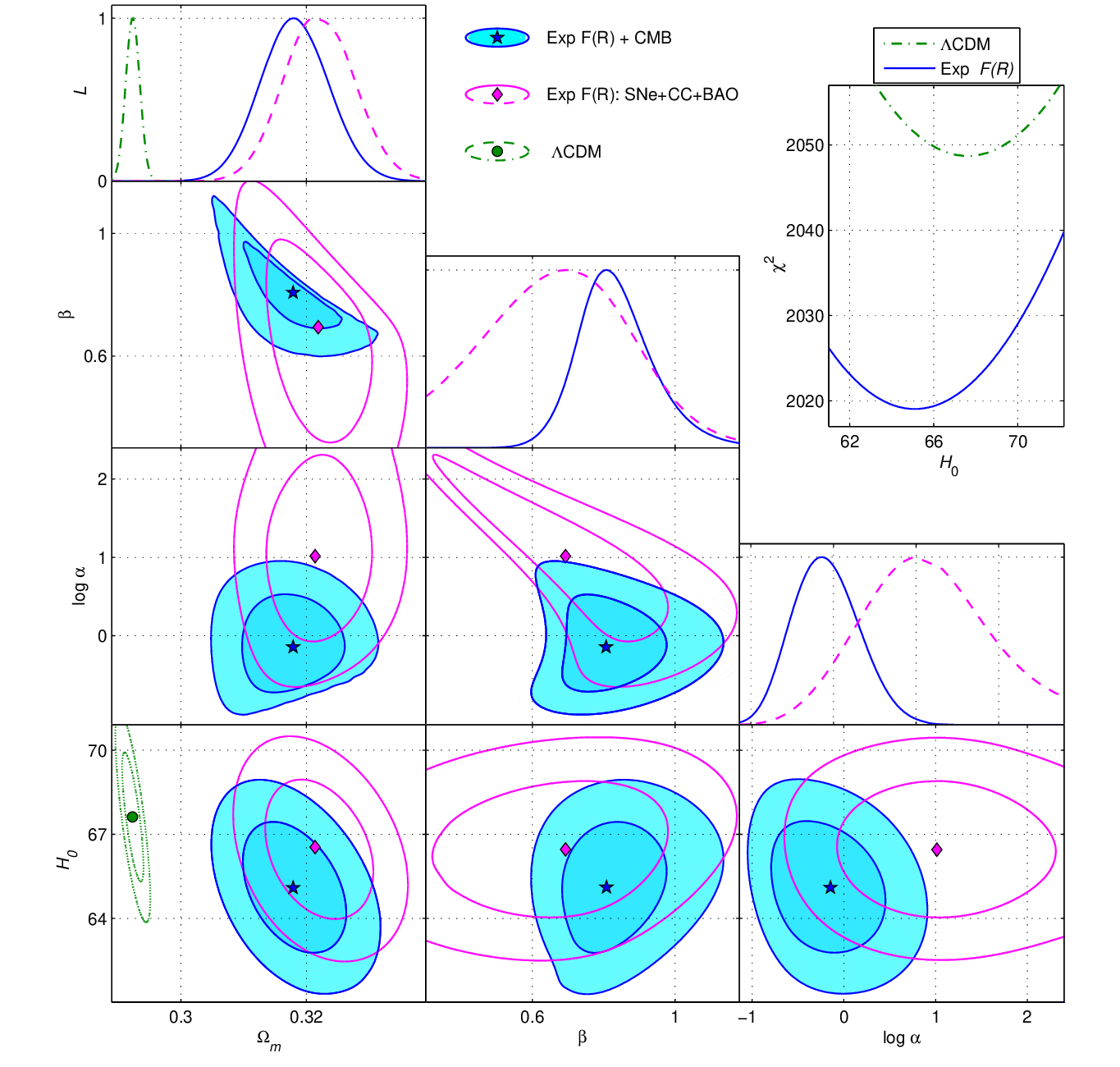}}
\caption{Contour plots of $\chi^2_\mathrm{tot}$, likelihoods ${\cal L}(\theta_i)$
distributions $\chi^2_\mathrm{tot}(H_0)$ for the generalized exponential $F(R)$ model
\eqref{GenExp} for SNe Ia, BAO DESI, $H(z)$ CC and CMB
data in comparison with $\Lambda$CDM model and the $\chi^2$ function \eqref{chitot}. }
  \label{F4}
\end{figure}

The generalized exponential $F(R)$ \eqref{GenExp} and $\Lambda$CDM
\eqref{LCDM} models appear to be more appropriate in confrontation
with all the  SNe Ia, BAO DESI, CC and CMB data. The results of
the test with  $\chi^2_\mathrm{tot}$ function  \eqref{chi4} are
presented in Table~\ref{Est2} and Fig.~\ref{F4}. These
calculations support the results of Ref.~\cite{OdintsovOS:2025},
but with the renewed approach for BAO DESI DR2 data
\cite{DESI:2025zgx} and 34 $H(z)$ CC datapoints from
Ref.~\cite{Pan:2025qwy}.
\begin{table}[ht]
\begin{center}
\caption{Best fits with $1\sigma$ errors, $\min\chi^2$, AIC,  from
SNe Ia, BAO DESI, $H(z)$ CC and CMB data, for the generalized
exponential $F(R)$ model \eqref{GenExp} and the $\Lambda$CDM model
(\ref{LCDM}).}
\begin{tabular}{|l|c|c|c|c|c|c|}  \hline
 \hline  Model &   $\min\chi^2/d.o.f$& AIC & $\Omega_m^0$& $H_0$&   other parameters \\
\hline
 Exp $F(R)$ & 2019.04 /1741 & 2029.04& $0.3179^{+0.0056}_{-0.0053}$ & $65.10^{+1.55}_{-1.54}$  & $\,\alpha=0.865_{-0.282}^{+0.480}$,
$\beta=0.807_{-0.082}^{+0.104}$, $\Omega_\Lambda=0.662_{-0.082}^{+0.073}$ \rule{0pt}{1.1em}  \\
\hline
$\Lambda$CDM& 2048.71 /1744 & 2052.71& $0.2923^{+0.0011}_{-0.0011}$& $67.62^{+1.52}_{-1.53}$ & -  \rule{0pt}{1.1em}  \\
\hline
  \hline \end{tabular}
 \end{center}
 \label{Est2}
\end{table}
From these calculations we can draw the main conclusion: if we add
the CMB observational data, the large advantage of the $F(R)$
scenario over the $\Lambda$CDM model is conserved in
$\min\chi^2_\mathrm{tot}$ and in AIC:
$\Delta\min\chi^2_\mathrm{tot}\simeq-29.67$ and
$\Delta\mathrm{AIC}\simeq-23.67$ in favor of the $F(R)$ model.
Hence, this advantage is  not connected with CMB, but with BAO
DESI along with SNe Ia observational data related to $z\in[0,3]$,
where in the $F(R)$ model the dark energy EoS parameter
$\omega_{DE}(z)$ \eqref{EoSDE} evolves from a phantom to a
quintessence stage.
The  bottom-left panel of  Fig.~\ref{F4} with the $\Omega_m^0-H_0$
plane demonstrates that the additional CMB data slightly
diminishes the best fits both for $\Omega_m^0$ and $H_0$ for both
models. But the most essential changes can be seen for the best
fits of $\alpha$: they change from $\alpha=2.76_{-1.42}^{+3.34}$
for $\chi^2$ to $\alpha=0.865_{-0.282}^{+0.48}$ for
$\chi^2_\mathrm{tot}$. To unify contour plots for both cases we
use the logarithmic scale $\alpha$ in the corresponding planes in
Fig.~\ref{F4}. The best fits for $\beta$ with the CMB data are
slightly enhanced, but the error boxes and $1\sigma$, $2\sigma$ CL
domains become noticeably more narrow. Note that the Lagrangian
\eqref{GenExp} of the generalized exponential $F(R)$ model in the
limit $\beta\to\infty$ tends to the $\Lambda$CDM Lagrangian
$F(R)=R-2\Lambda$. The best fitted values
$\beta=0.807_{-0.082}^{+0.104}$ are low, hence the considered
exponential $F(R)$ model works far from its $\Lambda$CDM limit.

\section{Conclusions}

In this article, we explored the $F(R,\phi)$ scenario
(\ref{actionmatter}), (\ref{FRpower}) originated from the
ghost-free non-local gravity in comparison with the generalized
exponential $F(R)$ model \eqref{GenExp} and the $\Lambda$CDM
model. To determine a late-time evolution of the $F(R,\phi)$
model, we solved the system of equations
\eqref{eqH1}\,--\,\eqref{eqPsi1} starting from the initial
conditions at the present time $t=t_0$ and integrating to the past
time direction. This feature is the essential advantage of this
scenario, because the mentioned  approach is not acceptable for
the $F(R)$ model \eqref{GenExp} and many others $F(R)$ scenarios
\cite{OdintsovSGS_game:2024,OdintsovOS:2025,OdintsovOS:2026},
where integral curves sharply diverge if we integrate into the
past time direction. These models require another approach.

Viable solutions of the $F(R,\phi)$ model were confronted with the
observational data including the Pantheon+ Type Ia supernovae (SNe
Ia), the DESI DR2 measurements of BAO and the Hubble parameter
estimations $H(z)$ (Cosmic Chronometers). The $F(R,\phi)$ scenario
appeared to be rather successful: its $\min\chi^2$ is comparable
with that of the generalized exponential $F(R)$ model and
essentially exceeds the $\Lambda$CDM result. This large advantage
is kept if we use the Akaike information criterion, taking into
account the number $N_p$ of free model parameters, though the
$F(R,\phi)$ model has the large number $N_p=7$ in contrast to
$N_p=2$ of the $\Lambda$CDM model.

However, if we add the CMB observed parameters to the mentioned
observational data, the $F(R,\phi)$ model encounters serious
difficulties. The total  $\chi^2_\mathrm{tot}$ function
\eqref{chi4} appears to be large, because the CMB data are related
with the recombination epoch at redshifts $z\sim1100$ and at these
redshifts this model has too large normalized Hubble rate $E(z)$
(shown in Fig.~\ref{F1}) for the best fitted solutions with
respect to SNe Ia, BAO DESI and CC data related to $z\in[0,3]$. We
can not bypass this difficulty if we start our calculations from
$z\sim1000$ epoch and the CMB data. So we may conclude that the
$F(R,\phi)$ model (\ref{actionmatter}), (\ref{FRpower}) needs some
modifications or correction terms to describe all the
observational data including CMB.

On the other hand, the generalized exponential $F(R)$ model
\eqref{GenExp} is very successful in tests with the whole set of
Pantheon+ SNe Ia, DESI BAO, $H(z)$ and CMB data. When we include
the CMB data, the best fits for its parameters $\alpha$ and
$\beta$ noticeably change as can be seen in  Fig.~\ref{F4}.
However, if we add the CMB observational data, the large advantage
of the $F(R)$ scenario over the $\Lambda$CDM model is conserved,
in particular, for SNe Ia, BAO and $H(z)$ data we observe the
difference $\Delta\min\chi^2\simeq-29.61$ between these models and
with CMB data the difference is
$\Delta\min\chi^2_\mathrm{tot}\simeq-29.67$ in favor of the $F(R)$
model. The same picture is seen if we compare AIC values.

From this fact we conclude that the large advantage of the
generalized exponential $F(R)$ scenario over the $\Lambda$CDM
model does not depend on CMB, but is connected with BAO DESI along
with SNe Ia observational data related to $z\in[0,3]$. One can see
in Fig.~\ref{F1} that the dark energy EoS parameter
$\omega_{DE}(z)$ \eqref{EoSDE} of the $F(R)$ model evolves from a
phantom to a quintessence stage at these redshifts. Such a form of
dynamical effective dark energy is observed in other successful
$F(R)$ scenarios
\cite{OdintsovSGS_game:2024,OdintsovOS:2025,OdintsovOS:2026},
whereas for the $F(R,\phi)$ model (\ref{actionmatter}),
(\ref{FRpower}) we see only a quintessence behavior with
$\omega_{DE}(z)>-1$. This conclusion is somewhat decisive that not
all modified gravity models can successfully be compatible with
all the data, and points out the elevated role of $F(R)$ gravity,
among all the modified gravities; it also signifies its
fundamental importance as a complete theory of gravity.

\section*{Acknowledgments}


\end{document}